\documentclass[12pt]{article}
 \usepackage{ifpdf}
 \usepackage{epsfig}
  \usepackage[latin1]{inputenc}
 \usepackage{amsfonts}
 \usepackage{amsthm}
 \usepackage{amssymb, latexsym, amsmath}
 \usepackage{amsbsy}
\usepackage{amstext}
\usepackage{amsopn}
\usepackage{amsgen}
\usepackage[dvips]{color}
 \usepackage{graphicx}
  \usepackage{color}
 \newcommand{\inc}{{\it i}}

 \newcommand{\be}{\begin{equation}}
 \newcommand{\ee}{\end{equation}}
 \newcommand{\ba}{\begin{eqnarray}}
 \newcommand{\ea}{\end{eqnarray}}

 \newcommand{\erbold}{\mbox{{\boldmath $\vec r$}}}

  \newcommand{\eRbold}{\mbox{{\boldmath $\vec{R}$}}}
  \newcommand{\Rbold}{\mbox{{\boldmath $\vec{R}$}}}

\begin{document}
  \title{
         ${{~~~~~~~~~~~~~~~~~~~~~~~~~~~}^{^{^{
         Published~in~the~
                 Astrophysical~Journal
        \,,~Vol.~764\,,~id.~26\,~(2013)
                  }}}}$\\
 {\Large{\textbf{{{Tidal Friction and Tidal Lagging. ~Applicability Limitations of a Popular Formula for the Tidal Torque}
  ~\\
  ~\\}
            }}}}
 \author{
  {\Large{Michael Efroimsky}}\\
 {\small{US Naval Observatory, Washington DC 20392}}\\
 {\small{e-mail: ~michael.efroimsky @ usno.navy.mil~}}\\
     ~\\
     {\Large{and}}\\
     ~\\
  {\Large{Valeri V. Makarov}}\\
  {\small{US Naval Observatory, Washington DC 20392}}\\
 {\small{e-mail: ~vvm @ usno.navy.mil~}}\\
 }
     \date{}

 \maketitle

  \begin{abstract}
 Tidal torques play a key role in rotational dynamics of celestial bodies. They govern these bodies' tidal despinning, and also participate in the subtle
 process of entrapment of these bodies into spin-orbit resonances. This makes tidal torques directly relevant to the studies of habitability of planets
 and their moons.

 Our work begins with an explanation of how friction and lagging should be built into the theory of bodily tides. Although much of this material can be found in various publications, a short but self-consistent summary on the topic has been lacking in the hitherto literature, and we are filling the gap.

 After these preparations, we address a popular concise formula for the tidal torque, which is often used in the literature, for planets or stars.
 We explain why the derivation of this expression, offered in the paper by Goldreich (1966; {\it{AJ}}~ {\bf{71}}, 1 - 7) and in the books by Kaula (1968, eqn.
 4.5.29), and Murray \& Dermott (1999, eqn. 4.159), implicitly sets the time lag to be frequency independent. Accordingly, the ensuing expression for the
 torque can be applied only to bodies having a very special (and very hypothetical) rheology which makes the time lag frequency independent, i.e, the same for all Fourier modes in the spectrum of tide. This expression for the torque should not be used for bodies of other rheologies. Specifically, the expression cannot be combined with an extra assertion of the geometric lag being constant, because at finite eccentricities the said assumption is incompatible with the constant-time-lag condition.
 \end{abstract}



 \section{Context, Motivation, and Plan
 }
   ${\left.~~~~~~\,~~~~~~~~~~~~~~~~~~~~~~~~~~~~~~~~~~~~~~~~~~~~~~~~~~~~~~~~~~~~~~~~~~~~~~~~~
   \,\right.}^{\mbox{\small \it The mills of God grind slowly...}}$
  ~~\\


 Usually extremely weak, tidal interactions act upon celestial bodies for extended spans of time (up to billions of years). Over {\ae}ons, tides shape celestial bodies' spin modes, and govern the exchange of angular momentum. The numerous and diverse
 manifestations of bodily tides range from the expected fall of Phobos
 to synchronous locking of the Moon,
 to Mercury's capture in the 3:2 spin-orbit resonance, to bloated hot-jupiter exoplanets in tight orbits around their host stars, to
 dissipational coalescence of short-period binary stars. This makes studies of tides essential to our understanding of the dynamical
 properties and evolution of stellar systems.

 While the slow work of tides is responsible for circularisation, obliquity evolution, and synchronisation of planets and moons, the wide scope
 of these dynamical phenomena is not always matched by sophistication or versatility of the tidal models employed to describe them.

 \subsection{Requirements to a consistent theory,\\
 and the history of simpler approaches}

 A {\it{bona fide}} theory of bodily tides implies (1) decomposition of the tide into Fourier harmonic modes and (2) endowment of each separate Fourier mode
 with a phase delay and a magnitude decrease of its own. The first part of this programme, Fourier decomposition, was accomplished in full by Kaula (1964),
 though a partial sum of the Fourier series was developed yet by Darwin (1879). The second part of this programme, the quest for an adequate frequency
 dependence of the phase lags and dynamical Love numbers, is now in progress. While earlier attempts seldom went beyond the Maxwell model, more realistic
 rheologies are now coming into use. A rheology combining the Andrade model at higher frequencies with the Maxwell model at the lowest frequencies was
 investigated by Efroimsky (2012a, 2012b).\footnote{~At lower frequencies, dissipation is predominantly viscous, and the mantle is well approximated with
 the Maxwell body. Its behaviour can be represented with a viscous damper and an elastic spring connected in series. Experiencing the same force, these
 elements have their elongations summed up. This illustrates how the total strain is comprised by a sum of viscous and elastic contributions
 generated by the same stress.

  At higher frequencies, the strain acquires the third component, one intended to describe inelasticity. Inelasticity is produced by defect
  unpinning, a process effective at frequencies higher than a certain threshold (about 1 yr$^{-1}$, in the case of Earth's mantle, -- see Karato
  \& Spetzler 1990). Hence, at frequencies above the threshold, dissipation is predominantly inelastic, and the mantle behaves as the Andrade body.

   A combined rheological model written down in Efroimsky (2012a, 2012b) embraces both frequency bands.

   In Makarov et al. (2012) and Efroimsky (2012a), we mistermed the Andrade creep as anelastic. It would be more accurate to call it {\it{inelastic}}, which means:
   irrecoverable. (The term {\it{anelastic}} is applied to recoverable deformations, like the Maxwell behaviour.)} The necessity for such a combined
 model originates from that fact that different physical mechanisms of friction dominate tidal dissipation at different frequencies.
 Several other rheological laws were probed by Henning et al. (2009) and Nimmo et al. (2012).

 Some authors tried to sidestep a Fourier decomposition by building simpler toy models which would preserve qualitative features of the consistent tidal
 theory and, ideally, would yield some reasonable quantitative estimates. Two radically simplistic {\it{ad hoc}} tools known as $\,${\it{ the constant
 geometric lag model}}$\,$ (MacDonald 1964, Goldreich 1966, Murray \& Dermott 1999) and $\,${\it{the constant time lag model}}$\,$ (Singer 1966; Mignard
 1979, 1980, 1981; Heller et al. 2011; Hut 1981) are often resorted to, and are applied to rocky moons and planets and gas giants and stars alike.
 Historically, both these approaches were introduced for the ease of analytical treatment rather than on sound physical principles. \footnote{~Aside
 from its mathematical simplicity, the constant time lag method is sometimes motivated by its analogy with the viscously damped harmonic oscillator. This
 analogy, however, appeared in the literature {\it{a posteriori}}, Alexander (1973) being the earliest work known to us where this analogy was spelled out.}

 \subsection{Plan}

 The first of the afore-said approaches, $\,${\it{the constant geometric lag model}}$\,$,
 will be addressed in this paper. Our goal is to demonstrate that the model should be discarded, both for physical and mathematical
 reasons. On the one hand, the model, is not well grounded in the physical reality, because it assumes a constant tidal response
 independent of the rotation frequency everywhere except at the 1:1 resonance where it singularly changes sign. On the other hand, the
 model is genuinely contradictive in its mathematics. The source of the inherent conflict is the popular formula for the tidal torque
 (and its analogue for the tidal potential) through which the model is implemented. It turns out that these formulae tacitly imply
 constancy (frequency-independence) of the time lag, a circumstance prohibiting the additional imposition of the constant-geometric-lag
 ( = frequency-independent quality factor) condition.

 Prior to executing the plan, we shall provide a comprehensive, review-style introduction into the methods of incorporation of friction
 into the tidal theory. The review will then enable us to recognise the afore-mentioned inconsistency in the the constant geometric lag model.

 In the subsequent publication (Makarov \& Efroimsky 2013), we shall explore the constant time lag model. This is an approach implying
 that all the tidal strain modes should experience the same temporal delay relative to the appropriate modes comprising the tidal stress.
 While the method is often assumed \footnote{~It can be demonstrated that the purely viscous model implies a frequency-independent time lag at sufficiently low
 frequencies only. Time lag acquires frequency dependence at frequencies higher than $\,G\rho^2R^2/\eta\,$, where $\,G$, $\,\rho$, $\,R$, and $\,\eta\,$ are the
 Newton gravity constant, mean density, radius, and the mean viscosity of the perturbed body. This circumstance lies outside the topic of this paper, and we shall
 elaborate on it elsewhere.} to work in the purely viscous limit (which, hypothetically, may be the case of stars and gaseous planets -- see Hut 1981
 and Eggleton et al. 1998), there is no justification for using it for terrestrial-type bodies such as the Moon, Phobos, or any exoplanet with
 a rocky or partially molten mantle. In the light of the current rheological knowledge, the tidal response is very different, and its
 frequency dependence experiences especially steep variations in the vicinity of spin-orbit resonances. In Makarov \& Efroimsky (2013),
 we shall demonstrate that illegitimate application of the constant time lag model to telluric objects leads to nonexistent phenomena
 like pseudosynchronous rotation -- not to mention that it squeezes the tidal-evolution timescales (Efroimsky \& Lainey 2007) and alters the
 probabilities of capture into resonances (Makarov, Berghea, \& Efroimsky 2012).


 \section{The constant-torque model}

 As we mentioned above, some authors tried to circumvent a consistent but laborious treatment, by building simpler toy models. One such attempt was
 undertaken by MacDonald
 (1964, eqn. 20) who assumed that the dynamical tide mimics a static tide, except for being displaced by a geometric lag angle. The same
 idea underlay the study, by Goldreich (1966), of a satellite on approach to the 1:1 spin-orbit resonance. Thus, for mathematical
 convenience, both authors set the geometric lag angle to be a frequency-independent constant.

 The approach is known as the {\it{constant angular lag model}}.
 %
 This name, however, is inexact in the sense that, within the vicinity of the 1:1 spin-orbit resonance, the lags and torque
 change
 their signs twice over a period: when the bulge falls behind or advances (relative to the direction towards the perturber), the sign is
 positive or negative, correspondingly. So in this discussion the term $\,${\it{constant}}$\,$ should be understood as $\,${\it{frequency-independent}}$\,$:
 both the instantaneous phase lag and the instantaneous torque $\,\vec{\cal{T}}
 \,$ are independent of the tidal frequency $\,\chi\,$. Consequently, the orbit-average (secular)
 tidal torque $\,\langle\,\vec{\cal{T}}
 \,\rangle\,$ is also frequency-independent.

  Sometimes this approach shows up in the literature under the name of
  {\it{MacDonald torque}} (e.g., Touma \& Wisdom 1994, section 2.7.1).

 Unfortunately, the approach turns out to be inconsistent and should be discarded. Physically, the constant angular lag model looks suspicious from the
 beginning, because in the vicinity of the 1:1 spin-orbit resonance it permits for abrupt switches of the torque, i.e., for situations where the torque
 changes its sign, retaining the absolute value. Although the abrupt switch can be substituted, by hand, with a continuous transition,
 this {\it{ad hoc}} alteration still would not save the method,
 because it would not heal a more fundamental defect. Mathematically, the derivation of the formula for the tidal torque within the said model contains a
 subtle and often unappreciated detail: this derivation implicitly sets the time lag $\,\Delta t\,$ to be constant, as we shall demonstrate below. However,
 it can be shown that the assertion of the time lag being frequency independent is incompatible with the assertion of the geometric lag angle being
 frequency independent. In this way, the discussed simplified approach is inherently contradictive.

 Another defect of this approach is that it employs such entities as the instantaneous phase lag and the instantaneous quality factor,
 the latter being introduced as the inverse sine of the former. The so-defined instantaneous quality factor is not guaranteed to
 be related to the energy damping rate in the same manner as the regular (appropriate to a fixed frequency) quality factor is
 related to the dissipation rate at that frequency (Williams \& Efroimsky 2012). Were the quality factor frequency-insensitive, this
 would not be a problem. However, the latter option is excluded within the discussed model,\footnote{~Rejection of the constant angular lag model does not, by itself, prohibit setting the quality factor constant, at least over some limited
 interval of frequencies. While realistic mantles never behave like this, such a rheology, in principle, is not impossible. However,
 employment of this rheology will not be available within a simplified model. Instead, one will have to attribute the same value to
 $\,k_l/Q_l\,$ at all tidal modes, and then will have to insert this value of $\,k_l/Q_l\,$ into all terms of the Fourier expansion of the
 tidal torque (the Darwin-Kaula series). Each such term will change its sign when the corresponding resonance gets transcended.

 Up to the late 60s of the past century, there was a consensus in the geophysical community that the seismic $\,Q\,$ of rocks should be
 ``substantially independent of frequency" (Knopoff 1964). This viewpoint was later disproved by a large volume of experimental data obtained
 both in the laboratory and in the field (see, e.g., Karato 2007 and references therein).} as being
 incompatible with the constant-$\Delta t\,$ assumption tacitly present.

 \section{Goldreich (1966), Kaula (1968),
 Murray \& Dermott (1999)}\label{sub}

 Numerous authors offer the following expression for the polar component of the torque wherewith the tide-raising perturber acts on the
 tidally-perturbed body:
 %
 %
 %
 \ba
 {\cal{T}}_z
 ~=~\frac{3}{2}~G\,M_{1}^{\,2}~\frac{R^5}{r^6}~k_2\,\sin2\epsilon_g~~,
 \label{1}
 \ea
 $R$ being the radius of the disturbed body, $\,M_{1}\,$ standing for the mass of the tide-raising perturber, $\,r\,$ denoting the
 instantaneous distance between the bodies, $\,\epsilon_g\,$ standing for the angular lag, and the obliquity being set nil. The
 subscript $\,z\,$ serves the purpose of emphasising that the above formula furnishes the torque component orthogonal to the equator of
 the tidally perturbed body.

 Goldreich (1966) denotes the angular lag with $\,\Psi\,$, $\,$Kaula (1968, eqn. 4.5.29) calls it $\,\delta\,$, $\,$while Murray \&
 Dermott (1999, eqn. 4.159) use the letter $\,\epsilon\,$. We shall follow the latter notation, though equipping it with the subscript
 $\,g\,$ which means: $\,${\it{geometric}}$\,$.

 Below we shall provide a detailed derivation of this formula, and shall see that the angle standing in it is indeed
 the instantaneous $\,${\it{geometric}}$\,$ tidal lag angle
 \ba
 \epsilon_g\,\equiv~(\dot{\nu}-\dot{\theta})\,\Delta t~\,,
 \label{}
 \ea
 i.e., the instantaneous angular separation between the direction towards the bulge and that towards the perturber. Here $\,\Delta t\,$ is the time lag,
 $\,\nu\,$ is the true anomaly of the perturber, $\,\theta\,$ is the sidereal angle of the perturbed body, and $\,\stackrel{\bf\centerdot}{\theta\;}$ is
 its spin rate. It is important to distinguish the instantaneous {\it{geometric}} lag $\,\epsilon_g\,$ from the instantaneous {\it{phase}} lag
 \ba
 \epsilon_{ph}\,\equiv~2\,(\dot{\nu}-\dot{\theta})\,\Delta t~=~2~\epsilon_g
 \label{}
 \ea
 sometimes used in the literature (Efroimsky \& Williams 2009, Williams \& Efroimsky 2012).

 In Section 5 of Murray \& Dermott (1999, eqns. 5.2 - 5.3), the authors rewrite the expression for the torque, employing
 (in fact, implying) the above expression of the lag through the true anomaly and spin rate:
 \ba
 \sin 2\epsilon_g\,=\,\sin|2\epsilon_g|~\mbox{Sgn}\left(\epsilon_g\right)\,=~-~\sin|2(\dot{\theta}\,-\,\dot{\nu})\,\Delta t|~
 \mbox{Sgn}(\,\dot{\theta}\,-\,\dot{\nu}\,)\,=~-~Q_s^{-1}~\mbox{Sgn}\left(\,\dot{\eta}\,-\,\dot{\varphi}\,\right)~\,.~
 \label{2}
 \ea
 The new angles showing up in this formula are $\,\eta\equiv\theta-{\cal{M}}\,$ and $\,\varphi\equiv\nu-{\cal{M}}\,$, with  $\,{\cal{M}}\,$ denoting the
 mean anomaly. These angles are depicted in Figure 5.1b in {\it{Ibid}}. Clearly, the time derivatives $\,\dot{\eta}=\dot{\theta}-n\,$ and $\,\dot{\varphi}
 =\dot{\nu}-n\,$ are the spin rate and true anomaly rate in a frame which is centered on the tidally perturbed body \footnote{~In Murray \& Dermott (1999),
 the role of a tidally perturbed body is played by the satellite, the planet acting as its tide-raising perturber. In a different setting, perturbed is the
 planet, the star or a satellite being the perturber.} and is rotating with the mean motion $\,n\,$. Interpreting the quantity $\,Q_s\,=\,1/\sin|2
 (\dot{\theta}\,-\,\dot{\nu})\,\Delta t|$ as the instantaneous quality factor, the authors assume that they can set it frequency-independent (thus making
 the torque frequency-independent).

 This approach contains two flaws. First, as explained in Williams \& Efroimsky (2012), it is not apparently evident whether the
 instantaneous quality factor introduced as the inverse sine of the instantaneous phase lag has the physical meaning usually instilled
 in a tidal dissipation factor at a certain sinusoidal mode. Second, and most important, is that in reality it is $\,\Delta t\,$ which
 gets implicitly set as frequency independent in the derivation of (\ref{1}). It then becomes impossible to assume that the geometric lag
 $\,\epsilon_g\,$ also is frequency-independent -- the two assumptions are incompatible, as we shall see shortly. Consequently, setting
 the factor $\,Q_s\,$ to be frequency independent is no longer an option. This makes the entire constant geometric lag model or, to be exact, its
 implementation with (\ref{1}), inherently contradictive. Specifically, it is illegitimate to assert that the tidal torque is proportional
 to Sgn$(\dot\nu-\dot\theta)\,$.

 \section{Mathematical Introduction.}

 The tide-raising potential $\,W\,$ created by a perturber always changes the shape and, as a result, the potential of the perturbed body. At
 the point $\,\eRbold\,$ of the perturbed body's surface, the potential $\,W(\eRbold,\,\erbold^{~*})\,$ created by the perturber residing at $\,\erbold^*\,$ can be
 expanded into a sum of terms $\,W_l(\eRbold,\,\erbold^{~*})\,$ proportional to the Legendre polynomials $\,P_{\it l}(\cos\gamma)\;$.
 Here $\,\gamma\,$ is the angle between the vectors $\,\erbold^{\,*}\,$ and $\,\eRbold\,$ pointing from the perturbed body's centre
 towards the perturber and the point on the perturbed body's surface, where the potential $\,W\,$ is measured.

 Tidal distortion of the body's geometric form renders an addition $\,U(\erbold)\,$ to the body's potential in an exterior point
 $\,\erbold\,$. This addition turns out to be comprised of terms $\,U_l(\erbold)\,$ each of which is proportional to the term
 $\,W_l(\eRbold,\,\erbold^{~*})\,$, with the surface point $\,\eRbold\,$ located exactly below (i.e., having the same latitude and
 longitude as) the exterior point $\,\erbold^{~*}\,$.

 The goal of this section is to provide a squeezed introduction into this formalism and to explain how it should be generalised from a
 static configuration setting to a dynamical setting.

 \subsection{Static tides}

 Let a body of radius $R$ experience tides from a perturber of mass $\,M^*_{1}\,$ placed at $\,{\erbold}^{\;*} = (r^*,\,\phi^*,\,
 \lambda^*)\,$, with $\,r^*\geq R\,$.

 At a point $\,\Rbold = (R,\phi,\lambda)\,$ on the perturbed body's surface, the potential $\,W\,$ due to the perturber is expanded over the Legendre polynomials
 $\,P_{\it l}(\cos\gamma)\;$ as \footnote{~Summation in formula (\ref{L1}) goes over $\,l\geq2\,$. The central term ($\,l=0\,$) is regarded as the principal,
 Newtonian, part of the potential generated by the perturber, and not as a part of the perturbation $\,W\,$ caused by the finite size of the tidally perturbed body
 -- indeed, the $\,l=0\,$ terms bears no dependence upon $\,\eRbold\,$. The reason why the $\,l=1\,$ terms falls out is more subtle and is related to the fact that
 we are developing our formalism in the frame of the tidally perturbed body, not in an inertial frame. See, e.g., Efroimsky \& Williams (2009, eqns. 5 - 11).}
 \ba
 \nonumber
 W(\eRbold\,,\,\erbold^{~*})&=&\sum_{{\it{l}}=2}^{\infty}~W_{\it{l}}(\eRbold\,,~\erbold^{~*})~=~-~\frac{G\;M^*_{1}}{r^{
 \,*}}~\sum_{{\it{l}}=2}^{\infty}\,\left(\,\frac{R}{r^{~*}}\,\right)^{\textstyle{^{\it{l}}}}\,P_{\it{l}}(\cos \gamma)~~~~\\
 \nonumber\\
 &=&-\,\frac{G~M^*_{1}}{r^{\,*}}\sum_{{\it{l}}=2}^{\infty}\left(\frac{R}{r^{~*}}\right)^{\textstyle{^{\it{l}}}}\sum_{m=0}^{\it l}
 \frac{({\it l}-m)!}{({\it l}+m)!}(2-\delta_{0m})P_{{\it{l}}m}(\sin\phi)P_{{\it{l}}m}(\sin\phi^*)~\cos m(\lambda-\lambda^*)~~,\quad\,
 \quad
 \label{L1}
 \ea
 where $\,\delta_{ij}\,$ is the Kronecker delta symbol, $\,G\,$ is Newton's gravity constant, while $\,\gamma\,$ denotes the angular
 separation between the vectors $\,{\erbold}^{\;*}\,$ and $\,\Rbold\,$ pointing from the centre of the perturbed body. The longitudes $\lambda,
 \,\lambda^*$ are reckoned from a fixed meridian on the perturbed body, the latitudes $\phi,\,\phi^*$ being reckoned from the equator.
 The integers $\,l\,$ and $\,m\,$ are called the {\it{degree}} and {\it{order}}, accordingly. The associated Legendre functions $\,P_{lm}
 (x)\,$ are referred to as the associated Legendre {\it{polynomials}} when their argument is sine or cosine of some angle.

 The $\,{\emph{l}}^{~th}\,$ term $\,W_{\it{l}}(\eRbold\,,~\erbold^{~*})\,$ of the perturber's potential introduces a distortion into
 the perturbed body's shape, assumed to be linear. Then the ensuing $\,{\emph{l}}^{~th}$ amendment $\,U_{\it{l}}\,$ to the perturbed
 body's potential will also be linear in $\,W_{\it{l}}\,$. Since $\,U_{\it{l}}(\erbold)\,$ falls off outside the body as
 $\,r^{-(\it{l}+1)}\,$, the overall change in the exterior potential of the perturbed body will be:
 \ba
 U(\erbold)~=~\sum_{{\it l}=2}^{\infty}~U_{\it{l}}(\erbold)~=~\sum_{{\it l}=2}^{\infty}~k_{\it
 l}\;\left(\,\frac{R}{r}\,\right)^{{\it l}+1}\;W_{\it{l}}(\eRbold\,,\;\erbold^{\;*})~~~,~~~~~~~
 ~~~~~~~~~~~~~~~~
 \label{L2}
 \ea
 where $\,k_l\,$ are the static Love numbers, $\,R\,$ is the mean equatorial radius of the perturbed body, while $\,\eRbold=(R,\,\phi,\,
 \lambda)\,$ and $\,\erbold=(r,\,\phi,\,\lambda)\,$ are a surface point and an exterior point above it, respectively, so that $\,r\geq R
 \,$.

 Combining (\ref{L2}) with (\ref{L1}), we arrive at a useful formula for the amendment to the potential of the tidally disturbed body:
 \ba
 U(\erbold)\;=\;\,-\,{G\;M^*_{1}}
 \sum_{{\it{l}}=2}^{\infty}k_{\it l}\;
 \frac{R^{
 \textstyle{^{2\it{l}+1}}}}{r^{
 \textstyle{^{\it{l}+1}}}{r^{\;*}}^{
 \textstyle{^{\it{l}+1}}}}\sum_{m=0}^{\it l}\frac{({\it l} - m)!
 }{({\it l} + m)!}(2-\delta_{0m})P_{{\it{l}}m}(\sin\phi)P_{{
 \it{l}}m}(\sin\phi^*)\;\cos m(\lambda-\lambda^*)~~.~~~
 \label{L3}
 \ea
 This is how the tidally generated change in the perturbed body's potential is ``felt" at a point $\,\erbold\,$. The change is expressed
 as a function of the spherical coordinates $\,\erbold=(r,\,\phi,\,\lambda)\,$ of this point and the spherical coordinates $\,\erbold^{
 \,*}=(r^*,\,\phi^*,\,\lambda^*)\,$ of the tide-raising body. The formula may be employed when we have two exterior bodies: if one
 such body, a perturber of mass $\,M^*_{1}\,$ located at $\,\erbold^{\,*}\,$ produces tides on the perturbed body, then the other
 exterior body, located at $\,\erbold\,$, will experience a potential perturbation (\ref{L3}) due to these tides.

 By changing variables from the spherical coordinates $\,\erbold^{\,*}=(r^*,\,\phi^*,\,\lambda^*)\,$ and $\,\erbold=(r,\,\phi,\,
 \lambda)\,$ to the Keplerian coordinates $\,\erbold^{\,*}=(\,a^*,\,e^*,\,\inc^*,\,\Omega^*,\,\omega^*,\,{\cal M}^*\,)\,$ and $\,
 \erbold=(\,a,\,e,\,\inc,\,\Omega,\,\omega,\,{\cal M}\,)\,$, one obtains a formula equivalent to (\ref{L3}):
 \ba
 \nonumber
 U(\erbold)\;=\;-\;\sum_{{\it
  l}=2}^{\infty}\;k_{\it l}\;\left(\,\frac{R}{a}\,\right)^{\textstyle{^{{\it
  l}+1}}}\frac{G\,M^*_{1}}{a^*}\;\left(\,\frac{R}{a^*}\,\right)^{\textstyle{^{\it
  l}}}\sum_{m=0}^{\it l}\;\frac{({\it l} - m)!}{({\it l} + m)!}\;
  \left(\,2\;\right. ~~~~~~~~~~~~~~~~~~~~~~~~~~~~~~~~~~~~~~~~~~~~~~~~~~\\
                                                                      \label{L4}\\
                                   \nonumber\\
                                    \nonumber
 ~~~~\left.-\,\delta_{0m}\right)\sum_{p=0}^{\it
  l}F_{{\it l}mp}(\inc^*)\sum_{q=-\infty}^{\infty}G_{{\it l}pq}
  (e^*) \sum_{h=0}^{\it l}F_{lmh}(\inc)\sum_{j=-\infty}^{\infty}
  G_{lhj}(e)\;\cos\left(\,
  \left(v_{{\it l}mpq}^*-m\theta^*\right)-
  \left(v_{{\it l}mhj}-m\theta\right)\,\right)
 \,~_{\textstyle{_{\textstyle ,}}}
 \ea
 where
 \ba
 v_{{\it l}mpq}^*\;\equiv\;({\it l}-2p)\,\omega^*\,+\,
 ({\it l}-2p+q){\cal M}^*\,+\,m\,\Omega^*~~~,
 \label{L5}
 \ea
  \ba
 v_{{\it l}mhj}\;\equiv\;({\it l}-2h)\,\omega\,+\,
 ({\it l}-2h+j){\cal M}\,+\,m\,\Omega~~~,
 \label{L6}
 \ea
 $\,q\,$ and $\,j\,$ being arbitrary integers, $\,p\,$ and $\,h\,$ beng arbitrary nonnegative integers, $\,F_{lmp}(\inc)\,$ being the inclination functions, while
 $\,G_{lpq}(e)$ being the eccentricity polynomials coinciding with the Hansen coefficients $\,X^{\textstyle{^{(-l-1),\,(l-2p)}}}_{\textstyle{_{(l-2p+q)}}}(e)\,$. Also mind that $\,\theta^*\,$ is the same as $\,\theta\,$, which is the sidereal angle of the tidally perturbed body. Following Kaula (1964), we equip
 $\,\theta\,$ with an asterisk, when it shows up in expressions corresponding to the tide-raising body. In expression (\ref{L4}), the terms $~-\,m\theta^*\,$ and
 $~-\,m\theta\,$ cancel one another, wherefore their presence may seem redundant. We better keep them, though, for they will no longer cancel when lagging
 comes into play.

 Decomposition (\ref{L4}) was pioneered by Kaula (1961, 1964). However, its partial sum, with $\,|{\it{l}}|,\,|q|,\,|j|\,\leq\,2\,$,
 was derived yet by Darwin (1879). In modern notations, Darwin's work is discussed by Ferraz-Mello, Rodr{\'{\i}}guez \& Hussmann
 (2008). \footnote{~Be mindful that the convention on the meaning of notations $\,\erbold\,$ and $\,\erbold^{\,*}\,$ in {\it{Ibid.}}
 is opposite to ours.}

 For our further developments, it would be important to emphasise that formulae (\ref{L3}) and (\ref{L4}) are equivalent to one another,
 because the latter is obtained from the former simply by a change of variables.

 \subsection{Dynamical tides with no friction}\label{fric}

 Derived for a static tide, formulae (\ref{L3}) and (\ref{L4}) extend trivially to an elastic dynamical setting where the tide
 adjusts instantaneously to the changing position $\,\erbold^{\,*}(t)\,$ of the perturber.

 The key point is that, to get $\,U(\erbold)\,$ at the point $\,\erbold\,$ at time $\,t\,$, we insert into (\ref{L3}) or (\ref{L4}) the perturber's position $\,\erbold^{\,*}(t)\,$ taken at that same time $\,t\,$, and not at an earlier time. Formulae (\ref{L3}) and (\ref{L4}) stay equivalent to one another, and remain
 unchanged, except that the distances, the sidereal angle, and the angular coordinates acquire a simultaneous time dependence:~\footnote{~The orbital parameters
 $\,a\,$, $\,e\,$, $\,i\,$, $\,a^*\,$, $\,e^*\,$, $\,i^*\,$ acquire no time dependence, insofar as the apsidal and nodal precession remain the only permitted
 variations of the orbits.} $\,r\,$ becomes $\,r(t)\,$, $\,r^{\,*}\,$ becomes $\,r^{\,*}(t)\,$; while $\,\theta\,$, $\,\theta^*\,$, $\,\phi\,$, $\,\phi^*\,$,
 $\,\lambda\,$, $\,\lambda^*\,$, $\,\omega\,$, $\,\omega^*\,$, $\,\Omega\,$, $\,\Omega^*\,$, $\,{\cal{M}}\,$, $\,{\cal{M}}^{\,*}\,$ become $\,\theta(t)\,$,
 $\,\theta^*(t)\,$, $\,\phi(t)\,$, $\,\phi^*(t)\,$, $\,\lambda(t)\,$, $\,\lambda^*(t)\,$, $\,\omega(t)\,$, $\,\omega^*(t)\,$, $\,\Omega(t)\,$, $\,\Omega^*(t)\,$,
 $\,{\cal{M}}(t)\,$, $\,{\cal{M}}^{\,*}(t)\,$.

 Thus, to obtain $\,U(\erbold(t)\,)\,$, we take the values of all variables at time $\,t\,$, leaving no place for any lagging. This is
 possible only for an absolutely elastic, i.e., frictionless perturbed body.

 \subsection{Tidal modes and forcing frequencies}

 Let us now write down the modes over which the tidal disturbance of the body gets expanded. We begin with expression (\ref{L1})
 for the perturbing potential at a fixed point $\,\eRbold=(R,\,\phi,\,\lambda)\,$ on the surface of the perturbed body. Using the technique
 developed by Kaula (1961, 1964), we change the coordinates of the tide-raising body from $\,\erbold^{\,*}=(r^*,\,\phi^*,\,
 \lambda^*)\,$ to $\,\erbold^{\,*}=(\,a^*,\,e^*,\,\inc^*,\,\Omega^*,\,\omega^*,\,{\cal M}^*\,)\,$. However, the location on the body's
 surface, where the disturbance is observed, is still parameterised with its spherical coordinates $\,\eRbold=(R,\,\phi,\,\lambda)~$:
 \ba
 \nonumber
  W(\eRbold\,,\;\erbold^{\;*})\;=\;-\;
  \frac{G\,M^*}{a^*}\;\sum_{{\it
  l}=2}^{\infty}\;\left(\,\frac{R}{a^*}\,\right)^{\textstyle{^{\it
  l}}}\sum_{m=0}^{\it l}\;\frac{({\it l} - m)!}{({\it l} + m)!}\;
  \left(\,2  \right. ~~~~~~~~~~~~~~~~~~~~~~~~~~~~~~~~~~~~~~~~~~~~~~~~~~~~~~~\\
                                   \nonumber\\
                                   \nonumber\\
       \left.
  ~~~ -\;\delta_{0m}\,\right)\;P_{{\it{l}}m}(\sin\phi)\;\sum_{p=0}^{\it
  l}\;F_{{\it l}mp}(\inc^*)\;\sum_{q=\,-\,\infty}^{\infty}\;G_{{\it l}pq}
  (e^*)
  \left\{
  \begin{array}{c}
   \cos   \\
   \sin
  \end{array}
  \right\}^{{\it l}\,-\,m\;\;
  \mbox{\small even}}_{{\it l}\,-\,m\;\;\mbox{\small odd}} \;\left(
  v_{{\it l}mpq}^*-m(\lambda+\theta^*)  \right)
 ~~~.~~~~~~~~~
 \label{L7}
 \ea
 Since $\,R\,$ is the radius of the tidally perturbed body, and since the latitude $\,\phi\,$ and the longitude $\,\lambda\,$ are reckoned
 from the equator and a fixed meridian, correspondingly, then (\ref{L7}) is just another expression for the perturbing potential at the
 fixed point $\,(R,\,\phi,\,\lambda)\,$ of the body's surface. In (\ref{L7}), the expression in round brackets can be reshaped as
 \ba
 v_{{\it l}mpq}^*-m(\lambda+\theta^*)\,=\,\omega_{{\it l}mpq}\,(t\,-\,t_0)~-~m~\lambda~+~v_{lmpq}^*(t_0)~-~m~\theta^*(t_0)~~,
 \label{L8}
 \ea
 where
 \ba
 \omega_{lmpq}\;\equiv\;({\it l}-2p)\;\dot{\omega}^*\,+\,({\it l}-
 2p+q)\;n^*\,+\,m\;(\dot{\Omega}^*\,-\,\dot{\theta}^*)\,~.~~~
 \label{L9}
 \ea
 Here $\,n^*\,\equiv\,{\bf{\dot{\cal{M}}}}^{\,*}\,$ is the mean motion of the perturber, while $\,t_0\,$ is the time of perigee
 passage wherefrom the mean anomaly $\,{\cal{M}}^{\,*}\,$ of the perturber is reckoned.

 We see from  (\ref{L8}) that the quantities $\,\omega_{lmpq}\,$ given by (\ref{L9}) are the Fourier modes over which
 the tidal perturbation (\ref{L7}) is expanded. While these modes can be positive or negative, the physical forcing frequencies,
 \ba
 \chi_{lmpq}~\equiv~|\,\omega_{lmpq}\,|\,~,~~~
 \label{L10}
 \ea
 at which the stress oscillates, are positive-definite.

 Having developed formulae (\ref{L4} - \ref{L7}), Kaula (1961, 1964) never stipulated \footnote{~The linear combination standing on the
 right-hand side of our formula (\ref{L9}) appeared in the denominator of formulae (29) and (50) in Kaula (1961). However, Kaula
 did not mention that this combination is a Fourier mode of the tide.} that the Fourier modes of the tide are given by
 (\ref{L9}). Possibly, he was not interested in the frequency dependence of the phase or time lags.

 In Section 6 of his book, Lambeck (1980) explained some aspects of Kaula's theory. While Lambeck's equation (6.1.13b) indicates that
 Lambeck could be aware of how the Fourier modes look, he too never wrote down the formula for the modes explicitly. Perhaps, like
 Kaula, Lambeck had no interest in the frequency dependence of lags -- he just introduced a time lag $\,\Delta t\,$, which in his
 developments was implicitly regarded frequency independent.

 While in the review by Efroimsky \& Williams (2009) and in Efroimsky (2012a, 2012b) the expression for $\,\omega_{lmpq}\,$ was written down
 explicitly, its derivation was omitted.

 Therefore, in the literature of which we are aware, the formula for the Fourier modes either was implied tacitly or was employed with
 no proof. This was our motivation to derive it here in such detail.

 To conclude, at the point $\,\eRbold=(R,\,\phi,\,\lambda)\,$ of the surface of the perturbed body, the perturbing potential is expressed through the tidal Fourier modes as:
  \ba
 \nonumber
  W(\eRbold\,,\;\erbold^{\;*})\;=\;-\;
  \frac{G\,M^*}{a^*}\;\sum_{{\it
  l}=2}^{\infty}\;\left(\,\frac{R}{a^*}\,\right)^{\textstyle{^{\it
  l}}}\sum_{m=0}^{\it l}\;\frac{({\it l} - m)!}{({\it l} + m)!}\;
  \left(\,2\;-\;\delta_{0m}\,\right)\;P_{{\it{l}}m}(\sin\phi)\;\sum_{p=0}^{\it
  l}\;F_{{\it l}mp}(\inc^*)~~~~~~~~~~\\
                                   \nonumber\\
                                   \nonumber\\
  ~~~ \sum_{q=\,-\,\infty}^{\infty}\;G_{{\it l}pq}
  (e^*)
  \left\{
  \begin{array}{c}
   \cos   \\
   \sin
  \end{array}
  \right\}^{{\it l}\,-\,m\;\;
  \mbox{\small even}}_{{\it l}\,-\,m\;\;\mbox{\small odd}} \;\left(\,
  \omega_{lmpq}\,(t\,-\,t_0)\,-\,m\,\lambda~+~v_{lmpq}^*(t_0)~-~m~\theta^*(t_0)\,\right)~\,~,~~~~~~
 \label{raz}
 \ea
 the tidal mode being given by (\ref{L9}).
 In an idealised situation, when the extended body is frictionless and its response is instantaneous, we can employ the static formula (\ref{L2}), as explained
 in subsection \ref{fric}. Combining that formula with expression (\ref{raz}), we see that the additional tidal potential generated by a perfectly elastic body
 at the point $\,\erbold=(r,\,\phi,\,\lambda)\,$ right above $\,\eRbold\,$ will now read as:
   \ba
 \nonumber
  U(\erbold\,,\;\erbold^{\;*})\;=\;-\;
  \frac{G\,M^*}{a^*}\;\sum_{{\it
  l}=2}^{\infty}\;\left(\,\frac{R}{r}\,\right)^{\textstyle{^{l+1}}}
  \left(\,\frac{R}{a^*}\,\right)^{\textstyle{^{\it
  l}}}\sum_{m=0}^{\it l}\;\frac{({\it l} - m)!}{({\it l} + m)!}\;
  \left(\,2 -\;\delta_{0m}\,\right)\;P_{{\it{l}}m}(\sin\phi)\;\sum_{p=0}^{\it
  l}\;F_{{\it l}mp}(\inc^*)~~~~\\
                                   \nonumber\\
                                   \nonumber\\
                                     \sum_{q=\,-\,\infty}^{\infty}\;G_{{\it l}pq}
  (e^*)~k_l~
  \left\{
  \begin{array}{c}
   \cos   \\
   \sin
  \end{array}
  \right\}^{{\it l}\,-\,m\;\;
  \mbox{\small even}}_{{\it l}\,-\,m\;\;\mbox{\small odd}} \;\left(\,
  \omega_{lmpq}\,(t\,-\,t_0)\,-\,m\,\lambda~+~v_{lmpq}^*(t_0)~-~m~\theta^*(t_0) \, \right)~~.~~~~
 \label{dva}
 \ea
 It is due to the lack of friction that the $\,lmpq\,$ term of (\ref{dva}) is in phase with the $\,lmpq\,$ term of (\ref{raz}). Below we shall see that inclusion
 of friction into the picture renders a phase shift between these terms. It is also in anticipation of the discussion of friction that we placed the Love numbers
 inside the $~\sum_{mpq}~$ sum in expression (\ref{dva}). Mode-independent in the perfectly elastic case, the Love numbers may acquire dependence upon
 the Fourier modes $\,\omega_{lmpq}\,$, because friction may mitigate amplitudes of distortion differently at different frequencies.

 \section{Friction and Lagging}

 In this section, we shall trace, step by step, how internal friction gets included into the tidal theory. In subsection \ref{fric}, we
 made an observation that, for an absolutely elastic (frictionless) body, treatment of dynamical tides mimics that of
 static tides, except that all coordinates acquire time-dependence. Our next step will be to incorporate friction, and therefore lagging,
 into the picture.

 As a first step, we shall address a simplistic method implying that the coordinates of the tide-raising body (as seen in a frame corotating with the perturbed body) get shifted back in time by
 some fixed time lag $\,\Delta t\,$. Although implementations of this method into formulae (\ref{L3}) and (\ref{L4}) look very different,
 they render results which in fact are equivalent -- simply because (\ref{L3}) and (\ref{L4}) are equivalent, and because the same procedure
 (shift by $\,\Delta t\,$) is performed on the quantities with asterisks in both these formulae.

 However, the difference in the mathematical form of these, equivalent, results also prompts a more consistent way of taking care of
 friction. This, more advanced, method will be implementable only in formula (\ref{L4}) and not in (\ref{L3}). The method is the one
 used by Kaula (1964). Since the explanation of the method in {\it{Ibid.}} was extremely concise, we shall elucidate it here in mode detail.

 \subsection{A naive way of bringing in lagging}

 Naively, dissipation and the ensuing lagging can be included into the picture by assuming that the exterior body located at point $\,\erbold\,$ at time
 $\,t\,$ is subject not to the tidal potential created simultaneously by the perturber residing at $\,\erbold^{\,*}(t)\,$, but to the potential generated
 by the perturber lagging in time on its orbit \footnote{~Here we imply: on its orbit {\it{as seen from the perturbed extended body}}. The caveat is
 needed, since we are considering the physical reaction of the extended body, and thus are interested in the location of the perturber relative to its
 surface, and not to an inertial frame. It is for this reason that in equation (\ref{L11}) we employ the latitudes and longitudes defined in a
 frame corotating with the perturbed extended body. In an inertial frame, a shifting of the perturber back by $\,\Delta t\,$ should then be accompanied by a shift of the orientation of the extended body back by the same $\,\Delta t\,$; and this is why we have $\,\theta^*(t-\Delta t)\,$ in equation (\ref{L12}).} by some $\,\Delta t\,$. Speaking loosely, the
 no-asterisk exterior body located at $\,\erbold(t)\,$ ``feels" the tide given by (\ref{L3}) or (\ref{L4}), as if the tide were generated
 by the asterisk perturber located on its orbit not at $\,\erbold^{\,*}(t)\,$ but at $\,\erbold^{\,*}(t\,-\,\Delta t)\,$.

 Mathematically, this implies that the tidal potential $\,U(\erbold)\,$ at $\,t\,$ must be calculated via (\ref{L3}) or (\ref{L4}), by
 insertion of the no-asterisk coordinates taken at $\,t\,$, and the coordinates with asterisk taken at $\,t\,-\,\Delta t\,$. The naive
 strategy also implies that the Love numbers $\,k_l\,$ keep their static values, though this detail is seldom spelled out.

 This approach implemented, our formulae (\ref{L3}) and (\ref{L4}) will acquire the following form:
 \ba
 \nonumber
 U(\,\erbold (t)\,)\,=&-&{G\;M^*_{1}}\sum_{{\it{l}}=2}^{\infty}k_{\it l}\;\frac{R^{
 \textstyle{^{2\it{l}+1}}}}{r(t)^{
 \textstyle{^{\it{l}+1}}}{r^{\;*}(t-\Delta t)}^{
 \textstyle{^{\it{l}+1}}}}\sum_{m=0}^{\it l}\frac{\left(l - m\right)!
 }{({\it l} + m)!}\left(2\right.\\
 \nonumber\\
 &\,&~\left.-\,\delta_{0m}\right)P_{{\it{l}}m}(\,\sin\phi(t)\,)P_{{
 \it{l}}m}\left(\,\sin\phi^*\left(t
 -\Delta t\right)\,\right)\;\cos m[\lambda(t)-\lambda^*(t-\Delta t)]~\,~.~\quad~\quad~
 \label{L11}
 \ea
 and
  \ba
 \nonumber
 U(\erbold)\;=\;-\;\sum_{l=2}^{\infty}\;k_{\it l}\;\left(\,\frac{R}{a}\,\right)^{\textstyle{^{{\it
  l}+1}}}\frac{G\,M^*_{1}}{a^*}\;\left(\,\frac{R}{a^*}\,\right)^{\textstyle{^l}}\sum_{m=0}^{\it l}\;
  \frac{({\it l} - m)!}{({\it l} + m)!}\;\left(\,2\;\right.~~~~~~~~~~~~~~~~~~~~~\\
                                                                                                         \nonumber\\
                                    \nonumber
 ~~~~\left.-\,\delta_{0m}\right)\sum_{p=0}^{l}F_{lmp}(\inc^*)\sum_{q=-\infty}^{\infty}G_{lpq}
  (e^*) \sum_{h=0}^{\it l}F_{lmh}(\inc)\sum_{j=-\infty}^{\infty}
  G_{lhj}(e)~~~~~~~\\
  \nonumber\\
  \cos\left(\,
  \left[v_{{\it l}mpq}^*(t-\Delta t)
 -\,m\theta^*(t-\Delta t)\,\right]
  - \left[v_{{\it l}mhj}(t)-m\theta(t)\,\right]\,\right)
 \,~_{\textstyle{_{\textstyle .}}}
 \label{L12}
 \ea

 Just as their static precursors (\ref{L3}) and (\ref{L4}), our dynamical formulae (\ref{L11}) and (\ref{L12}) remain equivalent to one another. They still reflect
 a mere switch from the spherical to the Keplerian coordinates, \footnote{~{\it{En route}}$\,$ from (\ref{L11}) to (\ref{L12}), one changes not only a coordinate system
 but also a frame of reference. Our longitude $\,\lambda\,$
 being reckoned from a meridian, the switch goes from the {\it{corotating}} coordinates $\,(r,\,\phi,\,\lambda)\,$
 to the Kepler coordinates $(a,\,e,\,i,\,\Omega,\,\omega,\,{\cal{M}})$
 defined in a frame comoving but not corotating with the perturbed body.

 Technically, one first substitutes $\,\lambda\,$
 with $\,\tilde{\lambda}-\theta\,$,
 where $\,\tilde{\lambda}=\lambda+\theta\,$
 is the longitude in the comoving (not corotating) frame. Then
 one can resort to the standard formulae connecting the spherical and Kepler coordinates in the same frame. The formulae apply not to $\,(r,\,\phi,\,\lambda)\,$
 but to  $\,(r,\,\phi,\,\tilde{\lambda})\,$,
 see Kaula (1961). Thus, the current spin rate $\,\theta(t)\,$ pops up in (\ref{L12}) due to the transition from a corotating frame to a comoving one.

 All said relates equally to both the spherical and Kepler coordinates with asterisks. So the delayed value $\,\theta^*(t-\Delta t)\,$, too, emerges in (\ref{L12})
 due to the frame switch. Recall that $\,\theta^*\,$ is the same spin rate as $\,\theta\,$, except that it gets equipped with an asterisk, when it stands in expressions corresponding to the perturber. Also recall that, within the described approach, we model friction by simply shifting the perturber (as seen in a frame corotating
 with the perturbed body) back in time by $\Delta t$. In a frame which is comoving but not corotating, this implies not only pulling the perturber back by $\,\Delta t\,$ but also rotating the perturbed body back by $\,\dot{\theta}\,\Delta t\,$.

 Leaving the coordinates $\,(r,\,\phi,\,{\lambda})\,$ untouched, and changing only $\,(r^{\,*},\,\phi^{\,*},\,\lambda^{\,*})\,$ to $\,(a^{\,*},\,e^{\,*}, \,i^{\,*},\,\Omega^{\,*},\,\omega^{\,*},\,{\cal{M}}^{\,*})\,$, one arrives at (\ref{L7}) and then at (\ref{raz} - \ref{dva}). Applying this machinery also to the variables with no asterisk, one ends up with (\ref{L12}).
 }
 except that now $\,\erbold\,$ is taken at time $\,t\,$, while $\,\erbold^{\,*}\,$ is taken at $\,t-\Delta t\,$.

 The longitude reckoned from a fixed meridian on the perturbed body is expressed through the true anomaly $\,\nu\,$, the periapse $\,
 \omega\,$ and the node $\,\Omega\,$ as
 \ba
 \nonumber
 \lambda\,=\,-\,\theta\,+\,\Omega\,+\,\omega\,+\,\nu\,+\,O(\inc^2)~\,.
 \ea
 In neglect of the nodal and the apsidal precession, and for a small obliquity, this results in
 \ba\nonumber
 \dot{\lambda}\,\approx\,-\,\dot{\theta}~+~\dot{\nu}\,+\,O(\inc^2)~\,.
 \ea
 In (\ref{L11}), the argument of cosine may now be written, in a linear approximation over $\,\Delta t\,$, as
 \ba
 m\,\left[\,\lambda(t)\,-\,\lambda^*(t-\Delta t)\,\right]\,=\,m\,\left[\,\lambda(t)\,-\,\lambda^*(t)\,+\,\dot{\lambda}^*\,\Delta t\,
 \right]\,=\,m\,\left(\,\lambda\,-\,\lambda^*\,\right)\,-\,m\,(\,\dot{\theta}\,-\,\dot{\nu}\,)\,\Delta t~+~O(\inc^2)~\,.~\,~\,
 \label{L13}
 \ea
 In (\ref{L12}), the argument of cosine may be shaped, in a linear approximation over $\,\Delta t\,$, as
 \ba
 \nonumber
 &~&\left[v_{{\it l}mpq}^*(t-\Delta t)-\,m\,\theta^*(t-\Delta t)\,\right]-\left[v_{{\it l}mhj}(t)-m\,\theta(t)\,\right]\,~=\\
 \label{L14}\\
 &~&~\quad~\quad~\quad~\quad~\quad~\left[v_{{\it l}mpq}^*(t)-\,m\,\theta^*(t)\,\right]-\left[v_{{\it l}mhj}(t)-m\,\theta(t)\,\right]\,~-~\epsilon_{lmpq}
 \nonumber
 \ea
 where
 \ba
 \nonumber
 \epsilon_{lmpq}&\equiv&\left[\dot{v}_{{\it l}mpq}^*(t)-\,m\,\dot{\theta}^*(t)\right]\,\Delta t\,=\,\left[({\it l}-2p)\;\dot{\omega}^*\,+\,({\it l}-
 2p+q)\;n^*\,+\,m\;(\dot{\Omega}^*\,-\,\dot{\theta}^*)\right]\,\Delta t\\
 \nonumber\\
 &=&\omega_{lmpq}~\Delta t
 \label{L15}
 \ea
 is the phase lag corresponding to the mode $\,\omega_{lmpq}\,$.

 From here we observe that the above-chosen method of taking the tidal friction into account fixes the phase lags in a very specific way: through shifting the
 perturber back on its orbit by a fixed time $\,\Delta t\,$, we set the phase lags (\ref{L15}) to be proportional to this $\,\Delta t\,$. It should be emphasised once
 again that the shift is performed in a frame corotating with the perturbed body. In a frame comoving but not corotating with it, the shift will thus be accompanied by
 rotation of the perturbed body back by $\,\dot{\theta}\Delta t\,$, hence the term $\,-\,\dot{\theta}\,\Delta t~$ in the expression $\,$(21)$\,$ for the phase lag.

 Our formulae (\ref{L11}) and (\ref{L12}) can be written in another, equivalent form:
 \ba
 \nonumber
  U(\erbold\,,\;\erbold^{\;*})\;=\;-\;
  \frac{G\,M^*}{a^*}\;\sum_{{\it
  l}=2}^{\infty}\;\left(\,\frac{R}{r}\,\right)^{\textstyle{^{l+1}}}
  \left(\,\frac{R}{a^*}\,\right)^{\textstyle{^{\it
  l}}}\sum_{m=0}^{\it l}\;\frac{({\it l} - m)!}{({\it l} + m)!}\;
  \left(\,2~-\;\delta_{0m}\,\right)\;   ~~~~~~~~~~~~~~~~~~~~~~~~~~~~~~~~~~~\\
                                   \nonumber\\
                                   \nonumber\\
  P_{{\it{l}}m}(\sin\phi)\;\sum_{p=0}^{\it
  l}\;F_{{\it l}mp}(\inc^*)\;\sum_{q=\,-\,\infty}^{\infty}\;G_{{\it l}pq}
  (e^*)~k_l~
  \left\{
  \begin{array}{c}
   \cos   \\
   \sin
  \end{array}
  \right\}^{{\it l}\,-\,m\;\;
  \mbox{\small even}}_{{\it l}\,-\,m\;\;\mbox{\small odd}} \;\left(\,
  \omega_{lmpq}\,(t\,-\,\Delta t\,-\,t_0)\,-\,m\,\lambda~+~v_{lmpq}^*(t_0)~-~m~\theta^*(t_0) \, \right)~~.~~~~
 \label{tri}
 \ea
 This form is analogous to (\ref{dva}), except for the time lag $\,\Delta t\,$, the same for each Fourier mode.

 Expression (\ref{tri}) is equivalent to expression (\ref{L11}), and is obtained from it by a switch from the spherical coordinates
 $\,\erbold^{\,*}=(r^*,\,\phi^*,\,\lambda^*)\,$ to the Kepler elements $\,\erbold^{\,*}=(\,a^*,\,e^*,\,\inc^*,\,\Omega^*,\,\omega^*,\,{\cal M}^*\,)\,$,
 with the variables $\,\erbold=(r,\,\phi,\,\lambda)\,$ kept unchanged. From the equivalence of expressions (\ref{tri}) and (\ref{L11}), we observe that,
 after the same time lag is added into all terms of (\ref{tri}), all the so-shifted terms add up to the equilibrium bulge geometrically displaced in such a
 way as if it were a static bulge generated by the perturber at a slightly different time. No other rheology can make this claim because, more generally, the
 lag in each term in the expansion would correspond to its own increment in $\,t\,$. This tells us that by setting $\,\Delta t\,$ frequency-independent we impose
 a highly restrictive rheological rule, obedience to which cannot be expected of realistic mantles.

 A more profound problem of this approach lies in the fact that it is illegitimate to introduce lags, keeping at the same time the Love numbers unchanged. Mitigation
 of the magnitude and lagging of the phase are inseparably connected, though the link becomes apparent only within a consistent approach based on the Fourier expansion
 of the tide and on employment of one or another rheological law. That law will then define both lagging in phase and reduction in magnitude.

 \subsection{A consistent way of bringing in lagging (Kaula 1964)}

 The above expression (\ref{L15}) for phase lags contains in itself an obvious hint on how a general-type rheology should be built into the tidal theory --
 to that end, one simply has to endow each mode $\,\omega_{lmpq}\,$ with a time lag $\,\Delta t_l (\omega_{lmpq})\,$ of its own. Another adjustment is the
 mode dependence of the Love numbers: $\,k_l\,=\,k_l(\omega_{lmpq})\,$. Expression (\ref{L12}) will now become
 \ba
 \nonumber
 U(\erbold)\;=\;-\;\sum_{l=2}^{\infty}\;\left(\,\frac{R}{a}\,\right)^{\textstyle{^{{\it
  l}+1}}}\frac{G\,M^*_{1}}{a^*}\;\left(\,\frac{R}{a^*}\,\right)^{\textstyle{^l}}\sum_{m=0}^{\it l}\;
  \frac{({\it l} - m)!}{({\it l} + m)!}\;\left(\,2\;\right.~~~~~~~~~~~~~~~~~~~~~\\
                                                                                                         \nonumber\\
                                    \nonumber
 ~~~~\left.-\,\delta_{0m}\right)\sum_{p=0}^{l}F_{lmp}(\inc^*)\sum_{q=-\infty}^{\infty}G_{lpq}
  (e^*) \sum_{h=0}^{\it l}F_{lmh}(\inc)\sum_{j=-\infty}^{\infty}
  G_{lhj}(e)~~~~~~~\\
  \nonumber\\
  k_{\it l}(\omega_{lmpq})\;\cos\left(\,
  \left[v_{{\it l}mpq}^*(t)-\,m\theta^*(t)\,\right]
  - \left[v_{{\it l}mhj}(t)-m\theta(t)\,\right]\,-\,\epsilon_{l}(\omega_{lmpq})\,\right)
 \,~_{\textstyle{_{\textstyle ,}}}~~~~
 \label{L16}
 \ea
 where
 \ba
 \epsilon_{l}(\omega_{lmpq})\,\equiv\,\omega_{lmpq}\,\Delta t_{l}(\omega_{lmpq})\,=\,\left[\,({\it l}-2p)\;\dot{\omega}^*\,+\,({\it l}-
 2p+q)\;n^*\,+\,m\;(\dot{\Omega}^*\,-\,\dot{\theta}^*)\,\right]\,\Delta t_{l}(\omega_{lmpq})~\,.~\,~\,~\,~
 \label{L17}
 \ea
 In (\ref{L16} - \ref{L17}), we prefer to denote the phase and time lags not as $\,\epsilon_{lmpq}\,$ and $\,\Delta t_{lmpq}\,$, but as $\,\epsilon_l(
 \omega_{lmpq})\,$ and $\,\Delta t_l(\omega_{lmpq})\,$. Indeed, their dependence on the indices $\,mpq\,$ is solely due to the argument $\,\omega_{lmpq}~$.
 However, we cannot strip $\,\epsilon\,$ or $\,\Delta t\,$ of the subscript $\,l\,$, because the functional form of the frequency dependence of the phase or
 time lag is different for different $\,l\,$s. The tidal friction is not the same as the seismic friction, and the degree $\,l\,$ affects the tidal damping
 rate.\footnote{~The difference between the tidal and seismic friction and, accordingly, the difference of dissipation at different $\,l\,$s is unimportant
 in small bodies, where only the rheology matters. However things change in large planets where self-gravitation becomes a crucial factor in tidal friction
 (Efroimsky 2012a).} Hence in formulae (\ref{L16} - \ref{L17}) we have $\,\epsilon_l\,$ and $\,\Delta t_l\,$, and not just $\,\epsilon\,$ or $\,\Delta t\,$.

 We would also write down the expression for the tidal potential in terms of the Keplerian elements of the perturber and the spherical coordinates of the point
 where this potential is observed:
 \ba
 \nonumber
  U(\erbold\,,\;\erbold^{\;*})\;=~~~~~~~~~~~~~~~~~~~~~~~~~~~~~~~~~~~~~~~~~~~~~~~~~~~~~~~~~~~~~~~~~~~~~~~~~~~~~~~~~~~~~~~~~~~~~~~~~~~~~~~~~~~~~~~~~~~\\
 \nonumber\\
 \nonumber\\
 \nonumber
    -\;
  \frac{G\,M^*}{a^*}\;\sum_{{\it
  l}=2}^{\infty}\;\left(\,\frac{R}{r}\,\right)^{\textstyle{^{l+1}}}
  \left(\,\frac{R}{a^*}\,\right)^{\textstyle{^{\it
  l}}}\sum_{m=0}^{\it l}\;\frac{({\it l} - m)!}{({\it l} + m)!}\;
  \left(\,2~-\;\delta_{0m}\,\right)\;P_{{\it{l}}m}(\sin\phi)~\sum_{p=0}^{\it
  l}\;F_{{\it l}mp}(\inc^*)\;\sum_{q=\,-\,\infty}^{\infty}\;G_{{\it l}pq}
  (e^*)~~~~~\\
                                   \nonumber\\
                                   \nonumber\\
  k_l(\omega_{lmpq})~ \left\{
  \begin{array}{c}
   \cos   \\
   \sin
  \end{array}
  \right\}^{{\it l}\,-\,m\;\;
  \mbox{\small even}}_{{\it l}\,-\,m\;\;\mbox{\small odd}} \;\left(\,
  \omega_{lmpq}\,[\,t\,-\,\Delta t_l(\omega_{lmpq})\,-\,t_0\,]\,-\,m\,\lambda~+~v_{lmpq}^*(t_0)~-~m~\theta^*(t_0) \, \right)~~.~~~~~~~~
 \label{chetyre}
 \ea
 This form is analogous to (\ref{dva}) and (\ref{tri}), except for two details. First, each mode $\,\omega_{lmpq}\,$ now has a time lag $\,\Delta t_l=\,\Delta t_l(\omega_{lmpq})\,$ of its own. Likewise, the dynamical Love number at each Fourier mode is a function of this mode: $\,k_l = \,k_l(\omega_{lmpq})\,$.

 The reason why we
 wrote down $\,U(\erbold\,,\;\erbold^{\;*})\,$ in the above form is that it immediately furnishes an expression for the $\,${\it{geometric}}$\,$ lag angle of an arbitrary
 $\,lmpq\,$ bulge:
 \ba
 \delta_{lmpq}\,=~\frac{\omega_{lmpq}}{m}~\Delta t_l(\omega_{lmpq})\,~.
 \label{bulge}
 \ea
 For example, the geometric lag angle of the principal, semidiurnal bulge is $\,\delta_{2200}=\,\frac{\textstyle \omega_{2200}}{\textstyle 2}~\Delta t_2\,=\,(n-\dot{\theta})\,\Delta t_2\,$, where the time lag is taken at the appropriate, semidiurnal mode: $\,\Delta t_2=\Delta t_2(\omega_{2200})\,$.

 It is customary to introduce the convention that the phase and time lags and the Love numbers are functions not of the tidal mode $\,
 \omega_{lmpq}\,$ but of the positively defined frequency $\,\chi_{lmpq}\,\equiv\,|\,\omega_{lmpq}\,|\,$. Simplifying some
 calculations, this convention makes it necessary to introduce, by hand, sign factors into the terms of the Fourier expansions
 of the tidal force and torque (Efroimsky 2012b).


 In practical applications, most important is the special case when the exterior body located at $\,\erbold\,$ coincides with the tide-raising perturber located at
 $\,\erbold^{~*}\,$. In this situation, the perturber is acting upon itself through the tide it creates on
 the perturbed body. Keeping the phase lags intact, and identifying $\,\erbold(t)\,$ with $\,\erbold^{~*}(t)\,$, we may be tempted to
 mis-assume that the expression with no asterisk, standing in the argument of the cosine in (\ref{L16}), compensates the expression with
 an asterisk, so the phase lag becomes all that is left. However, this would only furnish us the secular term of the tidal potential
 $\,U(\erbold)\,$ wherewith the perturber acts upon itself through the medium of the tidally perturbed body. This term is proportional to
 $\,\cos\epsilon_{lmpq}\,$. As the $\,p,\,q\,$ and $\,h,\,j\,$ are independent pairs of integers, there also will be contributions with
 $\,\{p,\,q\}\neq\{h,\,j\}\,$. These are the oscillating components of the tidal potential $\,U(\erbold)\,$. Although their time
 average is nil, they do contribute to the heat production, while the appropriate oscillating components of the
 tidal torque may influence free librations. The topic was first addressed in Efroimsky (2012b), and was revisited by Makarov et al. (2012) who explored whether the oscillating part of the torque can influence capture into spin-orbit resonances. It has turned out that, naturally, the oscillating part of the torque alters the outcome of a particular realisation of the capture scenario, but leaves the statistics unchanged.

 \section{Calculation of the tidal torque}

 We would begin with emphasising a key circumstance concerning the two forms of expansion of the tidal potential. In the absence of
 friction, these expansions, (\ref{L3}) and (\ref{L4}), were equivalent. Their amended versions, (\ref{L11}) and (\ref{L12}), remain
 equivalent in the presence of friction, provided the latter obeys a special (often unphysical) restriction that the time lag $\,\Delta
 t\,$ is the same at all frequencies. Beyond that threshold, the equivalence of the two expansions failed to extend. Since expression
 (\ref{L11}) does not contain the Fourier modes in it, it is plainly impossible to write (\ref{L11}) in a form that takes into account
 different time lags at different frequencies. Therefore, any calculation based on (\ref{L11}) will unavoidably imply the
 frequency-independence of $\,\Delta t\,$ and will thus be incompatible with any other rheological law.

 \subsection{Tidal torque, as derived from the concise expression (\ref{L11})}\label{5.1}

 Let us calculate the tidal torque, using (\ref{L11}). By employing this expression, we automatically set the rheology to be $\,\Delta
 t =\,$const~.

 Consider an exterior body of mass $\,M_1\,$ located at $\,\erbold\,$, which is subject to the additional tidal potential $\,U(\erbold)\,$
 of the tidally perturbed body. Then its energy in this potential will be $\,M_1\,U(\erbold)\,$. When the position of the exterior body
 is rendered by the spherical coordinates, the polar component of the torque acting on it can be conveniently expressed as: $~T_z\,=\,-\,M_{1}\;
 {\partial U(\erbold)}/{\partial\lambda}~$. The polar torque wherewith the exterior body acts back on the tidally
 perturbed body is the negative of $\,T_z\,$:
 \ba
 {\cal{T}}_z
 (\erbold)\;=\;\,M_{1}\;\frac{\partial U(\erbold)}{\partial\lambda}\,~.
 \label{L19}
 \ea
 Here {\it{polar}} means: orthogonal to the perturbed body's equator. For small obliquities
 (and, therefore, small latitudes), insertion of (\ref{L11}) into (\ref{L19}) yields
  \ba
 \nonumber
 {\cal{T}}_z
 (\,\erbold (t)\,)\,=&\,&{G\;M_{1}\;M^*_{1}}\sum_{{\it{l}}=2}^{\infty}k_{\it l}\;\frac{R^{
 \textstyle{^{2\it{l}+1}}}}{r(t)^{
 \textstyle{^{\it{l}+1}}}{r^{\;*}(t-\Delta t)}^{
 \textstyle{^{\it{l}+1}}}}\sum_{m=0}^{\it l}~m~\frac{\left(l - m\right)!
 }{({\it l} + m)!}\left(2\right.\\
 \nonumber\\
 &\,&~\left.-\,\delta_{0m}\right)P_{lm}(0)P_{lm}(0)\;\sin \left(\,m\,\left(\lambda-\lambda^*\right)\,-\,
 m\,(\dot{\theta}-\dot{\nu})\,\Delta t\,\right)~+~O(i^2)~\,~,~\quad~\quad~
 \label{L20}
 \ea
 where we made use of (\ref{L13}).

 Since the integer $\,m\,$ is now entering the above expression as a multiplier, the term with $\,m=0\,$ becomes nil. As $\,P_{21}(0)=0
 \,$, the term with $\,m=1\,$ also vanishes. Hence, the term with $\,l=m=2~$ is leading. Neglecting the smaller terms, we thus obtain:
 \ba
 {\cal{T}}_z(\,\erbold (t)\,)\,\approx\,\frac{\textstyle 3}{\textstyle 2}~{G\;M_{1}\;M^*_{1}}\,k_{2}\;\frac{R^5}{r(t)^{
 \textstyle{^{3}}}{r^{\;*}(t-\Delta t)}^{
 \textstyle{^{3}}}}\;\sin \left(\,2\,\left(\lambda-\lambda^*\right)\,-\,
 \,2\,(\dot{\theta}-\dot{\nu})\,\Delta t\,\right)~\,~.~\quad~\quad~
 \label{L21}
 \ea
 Consider the special case when the tide-raising perturber (with the asterisk) coincides with the other external body (with no
 asterisk). The perturber creates tides on the perturbed body, and then interacts with the tides it itself has created. Hence the
 perturber becomes subject to a tidal torque $\,\vec{\bf{T}}\,$ exerted by it upon itself, through the medium of the tidal bulge it
 creates on the perturbed body. Evidently, a torque $\,\vec{\cal{T}}\,=\,-\,\vec{\bf{T}}\,$, of the same magnitude but opposite direction, will
 be acting upon the perturbed body. We arrive at this torque by setting $\,\lambda=\lambda^*~$ and $~M_{1}=M^*_{1}~$ in the above expression:
 \footnote{~Be mindful that in (\ref{L22}) we chose to make no distinction between $\,r(t)\,$ and $\,r(t-\Delta t)\,$. Replacement of
 $\,r(t-\Delta t)\,$ with $\,r(t)\,$ gives birth to an absolute error of order $\,O(eQ^{-2}n/\chi)\,$. It is however explained in
 Efroimsky \& Williams (2009), that after averaging of (\ref{L22}) over one orbital period this error reduces to $\,O(e^2Q^{-3}n^2/\chi^2)
 \,$.}
 \ba
 {\cal{T}}_z(\,\erbold\,)\,\approx\,\frac{\textstyle 3}{\textstyle 2}~{G\;M_{1}^{\,2}}\,k_{2}\;\frac{R^5}{r^{
 \textstyle{^{6}}}
 }\;\sin \left(\,2\,(\dot{\nu}-\dot{\theta})\,\Delta t\,\right)~\,~.~\quad~\quad~
 \label{L22}
 \ea
 Naturally, for the model with a frequency-independent $\,\Delta t\,$, the quantity
 \ba
 \epsilon_g\,\equiv~(\dot{\nu}-\dot{\theta})~\Delta t
 \label{L23}
 \ea
 is the geometric lag, i.e., the angular separation between the planetocentric directions towards the perturber and the bulge.
 Accordingly, within the said model, the quantity
 \ba
 \chi~=~2\,|\,\dot{\nu}-\dot{\theta}\,|
 \label{L24}
 \ea
 acts as an {\it{instantaneous tidal frequency}}.

 The quantity
 \ba
 \epsilon_{ph}\,\equiv~2~(\dot{\nu}-\dot{\theta})~\Delta t~=~2~\epsilon_g
 \label{L25}
 \ea
 is commonly interpreted as an $\,${\it{instantaneous phase lag}}$\,$, so the torque in expression (\ref{L22}) may be written down as
 \ba
 {\cal{T}}_z(\,\erbold\,)\,\approx\,\frac{\textstyle 3}{\textstyle 2}~{G\;M_{1}^{\,2}}\,k_{2}\;\frac{R^5}{r^{
 \textstyle{^{6}}}}\;\sin\epsilon_{ph}~=~
 \frac{\textstyle 3}{\textstyle 2}~{G\;M_{1}^{\,2}}\,k_{2}\;\frac{R^5}{r^{
 \textstyle{^{6}}}}\;\sin2\epsilon_{g}
 ~\,~,~\quad~\quad~
 \label{L26}
 \ea
 which is exactly the expression (\ref{1}) of our concern. In Section 2, we mentioned several popular papers and books, including
 Murray \& Dermott (1999, eqn. 4.159), \footnote{~Mind a misprint in Eqn. (4.159) of Murray \& Dermott (1999): in the denominator, $\,a^6$
 must be changed to $\,r^6$. The misprint emerged because in subsection 4.2 the distance was denoted with $\,a\,$. In formulae (5.2 - 5.3) of
 {\it{Ibid.}} the misprint gets corrected.} $\,$where this formula is employed. Now, that we have derived this formula accurately, we see that its
 validity hinges on the time lag being frequency independent.

 While it is common (McDonald 1964, Goldreich 1966, Kaula 1968, Murray \& Dermott 1999) to treat
 \ba
 Q~\equiv~1/\,|\,\sin\left(\,2\,(\dot{\nu}-\dot{\theta})\,\Delta t\,\right)\,|~=~1/\sin |\,\epsilon_{ph}\,|
 \label{L27}
 \ea
 as an instantaneous quality factor, the validity of this interpretation of (\ref{L27}) remains questionable. For a nonzero eccentricity,
 the instantaneous tidal frequency (\ref{L24}) is varying in time. So it is not readily apparent whether the instantaneous $\,Q\,$ is
 connected to the damping rate in the manner the proper quality factor introduced at a certain frequency links to the damping rate at
 that frequency. McDonald (1964) and Goldreich (1966) tried to sidestep this difficulty by assuming that $\,Q\,$ is a frequency
 independent constant. However, at finite eccentricities this assumption does not work, as it is incompatible with the constant $\,\Delta t\,$ assumption
 (the latter assumption being a necessary prerequisite to using formulae \ref{L23} - \ref{L27}, as we saw above).
 For more on this see Williams \& Efroimsky (2012).

 \subsection{Tidal torque, as derived from the Fourier expansion (\ref{L12}),\\
 with all Fourier modes delayed by the same time lag $\,\Delta t\,$}

 When starting out with expression (\ref{L12}), it is convenient to use the formula
 \ba
 {\cal{T}}_z(\erbold)\;=\;-\;M_{1}\;\frac{\partial U(\erbold)}{\partial \theta}\;\;\;,
 \label{L28}
 \ea
 instead of (\ref{L19}). Technically, we should first differentiate $\,U\,$ with respect to the sidereal angle $\,\theta\,$, and then
 set $\,\theta\,=\,\theta^{\,*}\,$. We should also set the orbital elements with asterisk equal to their counterparts with no asterisk,
 it being understood that the tide-raising perturber is the same as the other exterior body which ``feels" the tides on the perturbed
 body. The development will furnish us the polar component of the torque with which the perturber acts upon the tidally deformed body.
 For exploration of dynamics in a low-obliquity configuration, this component is sufficient.

 Insertion of the Fourier series (\ref{L12}) into equation (\ref{L28}) yields a Fourier series for the polar component of the torque, which is
 presented in Efroimsky (2012b). Here we shall not repeat this long formula, but shall only make an important comment on it. Insofar
 as $\,\Delta t\,$ stays frequency independent (i.e., has the same value for all phase lags (\ref{L15}) entering the expansion for the
 torque), the resulting series for the torque stays fully equivalent to (\ref{L20}), with $\,\lambda\,$ and $\,\lambda^*\,$ set equal
 to one another in the latter formula.

 This equivalence is ensured by the expression (\ref{L12}) for the potential $\,U\,$ being equivalent to the expression (\ref{L11})
 whence formula (\ref{L20}) originated, and by our agreement to keep $\,\Delta t\,$ the same for all phase lags.

 \subsection{Tidal torque, as derived from the Fourier expansion (\ref{L16}),\\
 with each mode $\,\omega_{lmpq}\,$ having a time lag of its own, $\,\Delta t_l(\omega_{lmpq})\,$}

 As soon as we abandon the assumption that the phase lags (\ref{L15})
 contain the same fixed $\,\Delta t\,$, i.e., as soon as we switch from the
 lags (\ref{L15}) to those rendered by (\ref{L17}), we acquire an opportunity to describe a tidal torque acting on a perturbed body of
 an arbitrary rheology. Indeed, as the time lags can have an arbitrary mode-dependence, this also relates to the phase lags. Above that,
 we now permit the Love numbers to be mode-dependent.

 To derive the tidal torque, we now combine formula (\ref{L28}) not with expansion (\ref{L12}) but with the expansion (\ref{L16}) where
 the time lags are, generally, all different. The rheological emancipation, though, comes at a cost: the Fourier decomposition for the
 torque, obtained through (\ref{L16}), with mode-dependent $\,\Delta t_l(\omega_{\textstyle{_{lmpq}}})\,$ and $\,k_l(\omega_{\textstyle
 {_{lmpq}}})\,$, will no longer be equivalent to the concise and elegant formula (\ref{L20}). The customary and widely used leading-order
 approximation of (\ref{L20}), given by (\ref{L22}) or by (\ref{L26}), will not work for an arbitrary rheology.

 If, for example, we choose to set the factors $~\,\frac{\textstyle k_l(\omega_{\textstyle{_{lmpq}}})}{\textstyle Q_l(\omega_{\textstyle{_{lmpq}}})}~\mbox{Sgn}(\omega_{\textstyle{_{lmpq}}})\,=\,k_l\,\sin\epsilon_l~\,$ to be mode independent, then, to calculate the
 torque, we shall have to plug the same value of $\,k_l\,\sin\epsilon_l\,$ into all terms of the Fourier series for the torque (eqn. 106
 from Efroimsky 2012b). However, we shall not be able to employ the neat formula (\ref{L26}).

 \section{Why the $\,`\mathbf{\emph{constant angular lag}}\,${\it{'}}$\,$
 model is wrong}

 As we explained above, an accurate derivation of the popular formula (\ref{1}) for the polar component of the torque hinges on a tacit
 assumption that the time lag $\,\Delta t\,$ is the same for all modes in the expansion of the tidal potential. Through formula
 (\ref{L23}), this, mode-independent time lag is related to the geometric lag angle $\,\epsilon_{g}\,$ present in formula (\ref{1}).

 Relation (\ref{L23}) tells us that, since $\,\Delta t\,$ is the same for all the tidal modes involved, then the geometric lag angle $\,\epsilon_{g}\,$ cannot
 be treated as a fixed constant -- even if we add a caveat permitting $\,\epsilon_{g}\,$ to switch signs when the value of $\,\dot{\nu}\,$ transcends
 $\,\dot{\theta}\,$. With this caveat or not, keeping $\,|\,\epsilon_{g}\,|\,$ constant is impossible simply for the reason that (for a nonvanishing
 eccentricity) the quantity $\,\dot{\nu}\,$ oscillates in time.\footnote{~While derivation of (\ref{1}) absolutely requires $\,\Delta t\,$ to be the same for
 all tidal modes, it does {\it{not}} require $\,\Delta t\,$ to be fixed in time. Therefore, in theory, we can save the constant angular lag model by tuning
 the time dependence of $\,\Delta t\,$ in such a special way that $\,\epsilon_g\,$ in (\ref{L23}) stays constant in time. This however would require the
 dissipative properties of the mantle to be fine-tuned, simultaneously at all frequencies, to ensure that, first, the time lags at all frequencies evolve
 but remain equal to one another and, second, that $\,\Delta t\,(\dot{\nu}-\dot{\theta})\,$ stays constant in time. As the evolution rate  of $\,\dot{\nu}-\dot{\theta}\,$ is defined by the orbit, such fine tuning of rheology is unrealistic.} So a constant $\,\Delta t\,$ is incompatible with a constant $\,|\,\epsilon_{g}\,|\,$.

 This is the reason why the so-called $\,${\it{constant angular lag}}$\,$
 model based on (\ref{1}) must be discarded wholesale as being inherently contradictive.

 While we still retain the right to set the factors $~\,\frac{\textstyle k_l(\omega_{\textstyle{_{lmpq}}})}{\textstyle Q_l(\omega_{\textstyle{_{lmpq}}})}~\mbox{Sgn}(\omega_{\textstyle{_{lmpq}}})\,=\,k_l\,\sin\epsilon_l~\,$ mode independent, the value of the torque
 resulting from this assumption has to be calculated by insertion of these factors into the full Fourier expansion of the torque and not
 into (\ref{1}). We may as well use (\ref{1}), but only for a constant time lag, and not for a constant geometric lag angle.

 \section{Conclusions}

 We have reexamined the common formula (\ref{1}) for the tidal torque, a formula which is equivalent to the expressions given in Sections 4 and 5 of Murray \& Dermott (1999) and
 to the expressions offered in Goldreich (1966) and Kaula (1968). It has turned out that an accurate derivation of this popular formula necessarily implies a specific rheology --
 the assertion that the time lag $\,\Delta t\,$ is frequency independent. As can be easily seen from (\ref{L23}), this assertion is incompatible with the assertion of
 the geometric lag being frequency independent. Moreover, the quantity $\,\epsilon_g\,$ furnished by formula (\ref{L23}) can be endowed
 with the meaning of a geometric lag only within the constant $\,\Delta t\,$ rheological model (and only for a small obliquity $\,i\,$).

 To conclude, whenever the analysis of bodily tides is carried out using (\ref{1}), the analysis cannot be combined with a
 constant geometric lag (or phase lag, or quality factor) assumption, nor with any other assumption different from the
 frequency independence of $\,\Delta t\,$. This circumstance would not, by itself, prohibit one from considering a material for which the factors
 \footnote{~We deliberately equip the quality factors with the subscript $\,l\,$, to emphasise that they are different from the seismic
 $\,Q\,$ and have different frequency dependencies for different $\,l\,$s (Efroimsky 2012a).} $~\,\frac{\textstyle k_l(\omega_{\textstyle{_{lmpq}}})}{\textstyle Q_l(\omega_{\textstyle{_{lmpq}}})}~\mbox{Sgn}(\omega_{\textstyle{_{lmpq}}})\,=\,k_l\,\sin\epsilon_l~\,$
 are insensitive to the frequency {\it{over some limited frequency band}}. $\,$Consistent employment of this model will then require
 insertion of the same value of $~\,k_l\,\sin\epsilon_l
 ~\,$ into all terms of the expansion of the torque (each term thus changing its sign as the corresponding resonance is transcended). However, neither the
 quantity $\,(\dot\nu\,-\,\dot\theta)\,$ nor its sign will come into play in this expression for the torque. So the outcome will be
 different from the mathematically incorrect ``constant angular lag" model based on equation (\ref{1}).

 \section*{Acknowledgments}

 The authors are grateful to Stanton Peale, who refereed the paper and whose comments and recommendations were of great help. One of the authors (ME) is indebted
 to Sylvio Ferraz Mello and James G. Williams for numerous enlightening discussions on the theory of tides.

 \end{document}